\documentclass[prd,twocolumn,floatfix]{revtex4}

\usepackage{amsmath}  % needed for \tfrac, \bmatrix, etc.
\usepackage{amsfonts}  % needed for bold Greek, Fraktur, and blackboard bold
\usepackage{graphicx}   % needed for figures

\newcommand{\be}{\begin{eqnarray}}
\newcommand{\ee}{\end{eqnarray}}

\newcommand{\vf}{{\bf f}}
\newcommand{\vv}{{\bf v}}

\newcommand{\ffF}{{\sf F}}
\newcommand{\ffU}{{\sf v}}
\newcommand{\ffw}{{\sf w}}
\newcommand{\ftF}{\mathbb{F}}

\newcommand{\ddt}[1]{ \frac{{\rm d} #1}{{\rm d} t}}
\newcommand{\dby}[2]{\frac{{\rm d}#1}{{\rm d}#2}}

\begin{document}

\title{Reduced-order Abraham-Lorentz-Dirac equation and the consistency of classical electromagnetism}

\author{Andrew M. Steane}
\email{a.steane@physics.ox.ac.uk} % optional
\affiliation{Department of Atomic and Laser Physics, Clarendon Laboratory, Parks Road, Oxford OX1 3PU, England.}

\date{\today}

\begin{abstract}
It is widely believed that classical electromagnetism is either unphysical or inconsistent,
owing to pathological behavior when self-force and radiation reaction are non-negligible.
We argue that there is no inconsistency as long as it is recognized that certain types of
charge distribution are simply impossible, such as, for example, a point particle
with finite charge and finite inertia. This is owing to the fact that negative inertial mass
is an unphysical concept in classical physics. It remains
useful to obtain an equation of motion for small charged
objects that describes their motion to good approximation without requiring
knowledge of the charge distribution within the object. We give a simple method
to achieve this, leading to a reduced-order form of the 
Abraham-Lorentz-Dirac equation, essentially as proposed by Eliezer, Landau and Lifshitz,
and derived by Ford and O'Connell.
\end{abstract}

%{self-force, radiation reaction, hyperbolic motion}

\maketitle

\section{Introduction}

This paper address two separate but related issues. One issue 
is the correct understanding of pathological behavior exhibited by
certain equations of motion when self-force is non-negligible.
The second issue is the formulation of a valid equation of motion
for a small charged body
that does not exhibit pathological behavior. In both cases the aim of
the present paper is to provide arguments that are rigorous but simple,
for results that have previously been obtained by more sophisticated methods.

It is a well known feature of classical electromagnetism that any full treatment of the
motion of a charged object
includes some subtleties surrounding the issue of {\em self-force} and radiation reaction.
One can use the theory to obtain an equation that describes, approximately, the motion
of a rigid spherical shell of charge subject to an arbitrary force. This is the
Abraham-Lorentz-Dirac (ALD) equation\cite{64Nodvik,88Hnizdo,92Yaghjian,97Rohrlich,00RohrlichA,98Gupta}.
This equation, and others like it, is
problematic because it is higher than second order in the time derivatives of position,
which can lead to problems with causality, and because it has pathological 
or `runaway' solutions in which, for example, 
an object accelerates when no force is applied.

It is well established that the pathological cases can be ruled out by the following
strategy. First, one replaces the ALD equation by a better approximation
to the exact equation of motion of a charged spherical shell
(equation (3.7) of \cite{97Rohrlich}, called the `Caldirola equation'; see also
\cite{92Yaghjian,98Jackson,56Caldirola}), and then one insists that 
an entity of given charge and observed mass cannot have a radius below a certain
minimum. This both ensures the absence of pathological behavior
\cite{56Caldirola,83Schwinger,97Rohrlich,06Medina,10Griffiths}, and is physically
to be expected because it is the condition that the
observed mass must exceed the electromagnetic contribution (as we will
expound further in the following).
However there is continuing argument about what is the right way to interpret this situation
\cite{97Rohrlich,08Yaghjian,09Gralla,10Griffiths,98Jackson}, and there is resurgent interest
in this whole area because modern laser technology makes it possible to experimentally
investigate radiation reaction phenomena \cite{14Burton}.
We will here put forward the point of view, also espoused by 
several earlier authors, that, if gravitational effects are negligible, then {\em a point-like particle with finite
charge and finite observed mass is a strict impossibility in classical physics} \cite{09Gralla,48Bohm,61Erber,91Ford}).
More generally, we will argue that the issue is that
there cannot exist an entity whose inertial mass is negative, and one should not
model any physical system as if it was equivalent to one containing an entity
whose inertial mass is negative. We are not the first to espouse this point of view,
and indeed the fact that negative bare mass leads to instability has long
been recognized \cite{55Wildermuth}; for reviews see \cite{61Erber,04Spohn,82Pearle}.
We re-assert and expound it a little further.

If, in the absence of gravitational effects, there cannot be any such thing as a point particle with
finite charge and mass in classical physics, then there is no exact or correct equation of motion
for such an entity in flat spacetime. The search for such an equation, or
for approximations to it, must be abandoned. 

The combination of gravitational and electromagnetic self-energy was considered by 
Arnowitt {\em et al.} \cite{60ArnowittA,60ArnowittB}. They show that, for a certain
specific (and very small) charge to mass ratio, the divergent terms in the self-energy cancel, so that
the point limit can be taken. However such an approach does not allow a
point-like model of electrons, whose charge to mass ratio is much higher; this will be briefly examined
in section \ref{s.gravity}.

It remains useful to identify equations of motion for small charged entities.
One strategy to avoid the problems is adopt the ALD equation, but 
add further boundary conditions that are chosen in such a way
as to rule out the runaway solutions in practice\cite{38Dirac,90Rohrlich,98Jackson,04Spohn}.
Another is to develop new equations of motion \cite{91Ford,93Ford}, or to
replace the ALD equation by a related equation such as the one commonly called the
`Landau-Lifshitz equation' \cite{48Eliezer,71Landau,10Griffiths,11Bulanov,14Burton,12Zangwill}.
The Landau-Lifshitz equation is a reduced-order form of the ALD equation; this strategy 
is essentially the one independently described by Eliezer 
in 1948 \cite{48Eliezer} and arguably should bear his name, so we shall refer to 
it as Eliezer-Landau-Lifshitz (ELL). (The precise relationship between the Landau-Lifshitz
equation and the one proposed by Eliezer is presented in the appendix.) 

The relationship of the work of Ford and O'Connell to that of Eliezer and Landau
and Lifshitz, and other approaches, is discussed in \cite{14Burton,03OConnell}. Landau and Lifshitz
did not so much derive their equation as give arguments to suggest it was reasonable.
Eliezer (independently\footnote{I learned from V. Hnizdo that the Landau-Lifshiftz equation was in the 
1948 edition of their book (ready for printing in 1947), but not the 1941 Russian first 
edition. It therefore only slightly predates Eliezer's work, 
and the latter gave a different and independent argument for substantially the same conclusion.})
went somewhat further by explicitly treating an extended charge distribution,
and noting that there could possibly exist a distribution such that the equation of motion
takes the form he proposed (equation (\ref{Eli}) or (\ref{Eliw}), or, slightly more generally,
(\ref{myEli})). Ford and O'Connell (FO) \cite{91Ford,93Ford,12OConnell} derived the same 
equation, in the low-velocity limit, for a specific rigid charge distribution (one whose form factor is Lorentzian), 
treated in a dipole approximation which is valid at low frequencies. Within those approximations
their result is exact for that charge distribution, and since they explicitly derived it
the equation might fairly be attributed to them.  Their method also extends to general charge distributions.
In the following I shall refer to the Landau-Lifshitz
and Eliezer-Ford-O'Connell equations together as ``ELL/FO" since for a general charge
distribution and at arbitrary velocities and accelerations they are both approximate and
only differ at the next higher order in a series expansion in powers of the characteristic distance or time
(the width of the charge distribution or the time taken by light to traverse it). In any case,
the form recommended by Landau and Lifshitz can be understood as a deliberate
modification, merely for calculational convenience, of the equation of 
Eliezer and of Ford and O'Connell (see appendix). 

Ford and O'Connell developed their treatment  from first principles, starting from
quantum electrodynamics. They obtained a quantum Langevin equation, from which a classical
equation of motion can be derived. An important feature
of their work is to show that the radiation reaction force is accompanied by
fluctuations, which renders some other approaches such as Dirac's inadmissible, because
the use of the advanced Green function leads to a situation in which the fluctuations
drive an instability. The calculation of Gralla {\em et al.}, by contrast, remains  valid,
as does the one presented here in section \ref{s.ALD}.
Ford and O'Connell's approach can also be applied directly to the classical problem by the use
of Poisson brackets instead of Heisenberg equations of motion \cite{03OConnell}, therefore
it also offers insight into the purely classical problem, and this is what one is interested in
when assessing whether or not classical physics is internally consistent.

Recently there have been different approaches that explore the relationship between the ALD and ELL/FO
equations. Medina \cite{06Medina} treats the exact equation of motion formally by writing
down integral expressions for the fields obtained from potentials in the Coulomb gauge, and
then expanding the integrand in a power series, thus obtaining the ELL/FO equation.
Gralla {\em et al.} \cite{09Gralla} start out from energy-momentum conservation
for continuous distributions of charge, via the stress-energy tensor. They reproduce the ALD
equation and also obtain the ELL/FO equation by a suitably constructed series expansion while
avoiding unphysical cases such as infinite field energy. Griffiths {\em et al.} \cite{10Griffiths}
take as their starting point the Caldirola equation and 
obtain from it both the ALD and ELL/FO equations as equally legitimate approximations.
In the present work we justify the use of the ELL/FO equation
by a straightforward argument inspired by, but much simpler than,
the approach taken by Gralla {\em et al.}.

\section{Avoiding unphysical cases} \label{s.mass}

First we consider the problem of runaway solutions to equations of motion. We will
argue that there is no known case of a runaway prediction when the motion is treated
using standard electromagnetism (Maxwell's equations and the Lorentz force
equation), as long as the equations are handled correctly.

First we consider the ALD equation, and then we consider the exact equation of motion for
an extended charged body.

The ALD equation describes the motion, under an applied external force, of a charged
spherical shell, to first approximation. It can be written (taking $c=1$ and metric signature
$(-,+,+,+)$):
\be
\ffF_{\rm ext}  + \frac{2}{3}q^2\left( -\frac{\dot{\ffU}}{R} + \ddot{\ffU} - \dot{\ffU}^2 \ffU \right)
+ O(R)
 = m_0 \dot{\ffU}   \label{ALD}
\ee
where $\ffF_{\rm ext}$ is the applied four-force, $\ffU$ is the four-velocity of the shell,
the dot signifies ${\rm d}/{\rm d}\tau$, and $q,R,m_0$ are the charge, proper radius and bare mass of the shell
($m_0$ includes a contribution from the internal stress as well as the energy density of the shell).
It is customary
to move the $\dot{\ffU}/R$ term to the right hand side, so as to obtain
\be
\ffF_{\rm ext}  + \frac{2}{3}q^2\left( \ddot{\ffU} - \dot{\ffU}^2 \ffU \right)
+ O(R)
 = m \dot{\ffU}   \label{ALDm}
\ee
where $m = m_0 + (2/3)q^2/R$ is called the {\em observed mass}. 
The equation is obtained from a power series expansion
of the exact self-force. The neglected higher order terms are small as long as the worldline
does not curve significantly on the distance- or time-scale set by the size of the sphere, $R$.
If we neglect the $O(R)$ terms, and set $\ffF_{\rm ext}=0$, we have
\be
\frac{2}{3}q^2\left( \ddot{\ffU} - \dot{\ffU}^2 \ffU \right)  = m \dot{\ffU}.
\ee
This has the solution $\dot{\ffU}=0$ (as expected) and also the solution
\be
\ffU = \ffU(0) e^{\Gamma \tau}
\ee
where
\be
\Gamma = \frac{3 m}{4 q^2}.
\ee
This is the famous runaway solution of the ALD equation. It defies physical
sense because it suggests the sphere (and the field) can acquire
energy and momentum even when no external force acts. It has been
much discussed because it was argued that ALD equation should be exact in
the point particle limit, $R \rightarrow 0$. In fact that limit does not
make sense for finite $q$ and $m$, as we argue more fully below. In any case,
the solution only respects the approximations assumed by the ALD equation if
$\Gamma \ll 1/R$, which implies
\be
\frac{3 m R}{4q^2} \ll 1
\ee
hence
\be
\frac{m_0 + m_{\rm ed}}{m_{\rm ed}} \ll 1
\ee
where $m_{\rm ed} = (2/3) q^2/R$. However this condition cannot be satisfied
for non-negative $m_0$. Hence, if the bare mass is greater than or equal to zero, the
neglected higher-order terms are non-negligible for the worldline described by
the runaway solution. This means the calculation has broken down and the solution 
is merely an artifact of treating the exact equation of motion incorrectly.

Next, consider the exact relativistic equation of motion for a given small charged entity.
The spatial part takes the form 
\be
\vf_{\rm ext} + \vf_{\rm self} + \vf_{\rm P} = \ddt{} (\gamma m_{00} \vv)
\ee
where $m_{00}$ is the `completely bare' mass (i.e. before internal stresses are
accounted for), $\vf_{\rm self}$ is the electromagnetic
self-force, and $\vf_{\rm P}$ is the self-force owing to internal stresses
in the object (Poincar\'e stresses).
For clarity, we emphasize that $\vf_{\rm self}$ includes all contributions to the
force owing to the electromagnetic field sourced by the object in question.
The object should not be considered to be point-like when calculating this force,
and the inertial term in $\vf_{\rm self}$ must not be neglected.

First let us consider rectilinear motion, in order to understand what gives rise to runaway or
self-accelerating solutions in this case.

For the case of rectilinear motion, 
by bringing the Poincar\'e term to the right hand side, one can always write
\be
\vf_{\rm ext} + \vf_{\rm self} = \ddt{} (\gamma m_0 \vv)
\ee
by defining the `bare mass' $m_0$ appropriately.
$m_0$ may be time-dependent. 

In all cases that have ever been calculated, $\vf_{\rm self}$ is opposed 
to $\dot{\vv}$ in the instantaneous rest frame. This
being so, it is easy to see that if
one allows $m_0 \le 0$ then one can find a solution with $\dot{\vv} \ne 0$ when
$\vf_{\rm ext} = 0$, or, more generally, runaway solutions for a variety of forms
of $\vf_{\rm ext}$. The source of this unphysical prediction is not some problem
with classical electromagnetism, as many authors have suggested. It is
simply an error arising from an incorrect assumption that a physical system
can have negative inertial mass.

Caldirola's equation of motion is an approximation to the exact equation of motion that
includes more terms than the ALD equation, and therefore may be expected
to give more accurate predictions \cite{56Caldirola,97Rohrlich,90Rohrlich,06Medina,92Yaghjian}.
Caldirola's equation of motion has no runaway solutions
when $m_0 > 0$. This is well established, and in view of what we have just said about
the exact equation of motion, it is wholly unsurprising. However 
the condition continues to be regarded as surprising because it is often stated
as a condition on the size rather than the mass of the body.
For example, when we treat a spherical shell of charge $q$ the
constraint can be written $R > (2/3)q^2/m c^2$, where $m$ is observed
mass, and Griffiths
{\em et al.} \cite{10Griffiths} comment ``it does seem peculiar that classical
electrodynamics should harbor such a counterintuitive constraint on its validity".
We argue that, on the contrary, the constraint is not counterintuitive and its ultimate
cause has nothing to do with electrodynamics {\em per se}.

\subsection{Illustrative example}

\begin{figure}
\begin{picture}(100,40)(0,0)
\thicklines
\put(0,20){\circle{15} \put(-30,-12){$m_1$} }
\put(80,20){\circle{15} \put(-10,-12){$m_2$} }
\thinlines
\put(74,10){ \line(-1,0){70}\line(0,1){20} \put(-0,20){\line(1,0){70}}}
\put(67,10){ \line(0,1){20} \put(2,0){\line(0,1){20}} }
\put(19,20){ \vector(-1,0){14}  \put(7,-6){$p_1$}}
\put(53,20){ \vector(1,0){14} \put(-14,-6){$p_2$}}
\end{picture}
\caption{Illustrative system consisting of two masses separated by a fluid-filled tube.}
\label{fig1}
\end{figure}
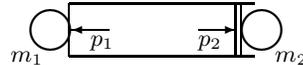

Figure \ref{fig1} shows a simple physical system that we shall use to illustrate
the use and abuse of inertial mass in classical physics. The system consists
of two small bodies of rest mass $m_1,\, m_2$, separated by a tube
containing an ideal fluid of pressure $p$. The tube and the fluid have negligible energy density. 
The whole system is uncharged; there are no electromagnetic fields in the problem.
We suppose this system to be undergoing motion at constant proper acceleration
(`hyperbolic motion') in the direction parallel to the tube. We assume the motion is rigid,
that is, it maintains the proper dimensions of the system. In
the instantaneous rest frame, the equations of motion for the two bodies are
\be
m_1 a_1 &=& f_1 - A p_1,    \nonumber \\ 
m_2 a_2 &=& f_2 + A p_2,                      \label{eqm12}
\ee
where $f_{1,2}$ are the external forces, $p_{1,2}$ are the values of the
pressure at the two ends of the tube, and $A$ is the cross-sectional area of the latter.

To find the pressures, we solve the relativistic Navier-Stokes equation, see \cite{14SteaneA},
which gives
\be
\frac{p_1}{p_2} = \frac{a_1}{a_2}.
\ee
Substituting this into (\ref{eqm12}), we find that for $f_1 = f_2 = 0$, i.e. no external
force, the system can nevertheless have $a_1 > 0$ and $a_2 > 0$, i.e. self-acceleration, if
and only if
\be
m_1 = - m_2.
\ee
Thus, here we have an example of a classical system showing self-acceleration in the absence
of any electromagnetic fields---if we admit the possibility of negative rest mass for part of
the system. Note, the system as a whole does not have zero mass when $m_1 = - m_2$,
because the pressure contributes an inertial mass of $p V / c^2$ where $V$ is the volume of
fluid in the tube. Thus the system might appear to a naive examination to be `possible'
when $p > 0$. In fact it is not possible.

Returning now to the case of a charged sphere, 
we are arguing that $m_0 > 0$ because this is simply one of the assumptions
of classical physics: inertia always opposes acceleration. However, in the case of a system
permanently in possession of its electric charge, the `bare' system with its bare mass is
just a convenient fiction introduced for the purposes of writing down an equation of motion.
One might want to suggest that for such systems the observed mass 
could in principle take on any positive value. This is exactly what has been
done in many discussions of runaway solutions to the equation of motion.
However those runaway solutions are themselves `giving the game away': they are
themselves a signal that an unphysical assumption has been smuggled in. The concept
of bare mass is a helpful way to clarify what that assumption is. 

Let $m_{\rm ed}$ be the contribution of the field to the observed mass
of the system `charged object plus its own field'.
(In the case of a spherical shell, $m_{\rm ed} = (2/3) q^2/R c^2$).
The assumption $m < m_{\rm ed}$ amounts to assuming
that the system in question is equivalent to one in which an electromagnetic field
surrounds and interacts with an entity whose momentum opposes its velocity and whose
kinetic energy is negative (because the mass of the entity is $m - m_{\rm ed}$). But no such
entity exists or could exist. One should not treat really existing charged particles such as
electrons as if they were equivalent to systems with unphysical internal properties such
as negative inertial mass.

One might want to suggest that, in the point-particle limit, somehow the electromagnetic
energy and the binding energy could result in $m < m_{\rm ed}$.
It might be proposed, for example, that $m < m_{\rm ed}$
could in some situations account correctly for the inertial effects when they
are averaged over a small region containing a charged body that is
stabilized by its own Poincar\'{e} stresses. However, this is not
possible, because for each shell of charge the electromagnetic part of the
energy resides in the electromagnetic field
{\em outside} the shell, so the only way to get $m < m_{\rm ed}$
through effects inside a shell is to have a net
{\em negative} rest energy inside the shell, and this is not allowed.
Another strategy to retain the concept of point-like charges
would be to suggest that there is some further physical effect around
the particle that lowers the total energy density there. This could be
valid (it is indeed exactly what happens, according to quantum field theory)
but one would then have to calculate the influence of this further effect
on the self-force.

Another way to modify the vacuum outside the charged body is to
allow for its gravitational field; this is considered in the next section.

Of course, all physical systems are `impossible' as far as classical mechanics is concerned,
since they are described by quantum not classical theory. However, the classical
limit of quantum theory should be expected to consist in physically
reasonable behavior and should be
treated accordingly \cite{77Moniz}. We simply have to accept that {\em a point-like particle with finite
charge and finite observed mass is a strict impossibility in classical physics} (with the 
exception of a special case of charge-mass ratio given in Eq. (\ref{qoverm})).
It is mathematically dubious and physically impossible
(others agree \cite{61Erber,91Ford,09Gralla}). When we call an electron
a `charged particle' we should keep in mind that that terminology can be misleading
and in fact is misleading in the case of motion where self-force is significant. The
excitations of the Dirac field do not behave like charged point-like classical particles,
not even in the classical limit, and it is only a muddle in understanding that limit
which lead people
to mistakenly think that they might. They do behave, to good approximation, 
like charged rigid spheres of finite radius, as long as there is no significant
structure in the applied fields and dynamics on the length scale (or associated time scale)
of the sphere's size. 
When those conditions are satisfied,
there is no need to be specific about the size, as long as it is above
the critical value, because it merely affects the way the observed mass is `parceled out'
between two contributions  ($m_0$ and $m_{\rm ed}$) that are themselves 
unobservable. The latter are `hidden variables' of a classical model of a charged
wave-packet; they have served their
purpose as soon as they have revealed the critical length scale. The critical length
scale is the length scale where $m = m_{\rm ed}$.

The rigid spherical shell is one allowable approximate model of an electron, but of course
it is not the only option. One may instead modify the Maxwell equations themselves \cite{99Frenkel,01Blinder}. One
physically motivated option is to treat the electron
as a point particle but modify the treatment of the vacuum, by allowing that it
can be polarized, see for example Blinder \cite{01Blinder}.

\subsection{Gravitational effects}  \label{s.gravity}

In the point limit the gravitational as well as the electromagnetic field diverges,
so that a thorough investigation of that limit must take account of both. 
Using a simple Newtonian model, the gravitational energy liberated by bringing together material of mass
$m$ from initially infinite separation to form a spherical shell of radius $R$ is $G m^2 / 2 R$.
Therefore one expects the total self-energy, including electrostatic and gravitational contributions, to be
\be
\frac{1}{2 R} \left( \frac{Q^2}{4 \pi \epsilon_0} - G m^2 \right)
\ee
where $Q$ is the charge and for the purpose of this equation we have adopted SI 
units (in Gaussian units the first term in the bracket is $q^2$).
This self-energy vanishes when
\be
\frac{Q}{m} = \sqrt{ 4 \pi \epsilon_0 G } \simeq 8.618 \times 10^{-11} \;{\rm C\, kg}^{-1}.  \label{qoverm}
\ee
So, for this specific value of $Q/m$, one may take the point limit $R \rightarrow 0$ without meeting
a divergence. However, this only rescues the theory for a particle having the special value of $Q/m$,
so the point limit remains highly questionable in general. It certainly cannot be used
as a model for electrons, whose charge/mass ratio is $2 \times 10^{21}$ times larger than the
special value.

Of course the above can only be regarded as an order-of-magnitude estimate, but a rigorous treatment by
Arnowitt {\em et al.} \cite{60ArnowittA,60ArnowittB} comes to the conclusion that a stable and well-behaved 
solution of the field equations of general relativity is possible for a charged shell in the point limit, 
with the ratio of charge to observed mass given by (\ref{qoverm}).

\section{Obtaining the ELL/FO equation}  \label{s.ALD}

We now turn to the issue of obtaining an equation of motion for a small
charged object.

Although the point charge (with finite charge but no extension) is
a useful tool for pedagogic purposes when students are first introduced to
electromagnetism, it must be dropped if we want to formulate a
mathematically consistent and physically sensible treatment of the subject. Instead
one may treat continuous distributions of charge with no impossible properties
such as divergent total energy. Point particles can still be treated as long as
they have vanishing charge and mass, as we illustrate below. 

The idea is to obtain a general equation of motion for a small charged body that, we 
accept, will be approximate, but whose approximation is satisfactory. That is, it
predicts, in an appropriate limit, motion close to  the
exact solution. In particular, it does not predict runaway behavior in the absence
of an applied force, and it is second order (in derivatives of position with respect to time)
so respects causality. To achieve this we shall
keep the degree of approximation in view, and thus arrive at a physically
sensible equation that can be written in terms of observable quantities such
as observed mass, without requiring information about the size of the object
(except the reassurance that it is small enough for the approximations to be valid).
Such an equation can then be used to treat the motion of entities such as electrons
to good approximation.

There is more than one possible strategy to 
achieve our end. Here we display a strategy inspired by
Gralla {\em et al.} but involving only very simple ideas.

First we write down the equation of motion of a rigid spherical shell of charge $q$
and bare mass $m_0$:
\be
\ffF_{\rm ext}  + \ffF_{\rm self} = m_0 \dot{\ffU}
\ee
where $\ffF_{\rm self}$ is given by the ALD Eq. (\ref{ALD}):
\be
\ffF_{\rm self} =  \frac{2}{3}q^2\left( -\frac{\dot{\ffU}}{R} + \ddot{\ffU} - \dot{\ffU}^2 \ffU \right)
+ O(q^2 R).
\label{ALDself}
\ee
If one wants higher accuracy one can expand the self-force
to higher order; see for example \cite{64Nodvik,13Steane}. In any case, as long as the equation of motion
is itself only correct up to some order in $R$, there is no need to find exact solutions.
For whatever order in $R$ one
has obtained an expression for $\ffF_{\rm self}$, it suffices 
to find solutions for $\ffU(\tau)$ that are accurate to that order. To this end, we will obtain the solution as
a power series in $R$, but taking care to express the limit $R \rightarrow 0$ in a mathematically
and physically sensible manner. This can be done by treating the shell as an example, at
given $R$, of an object whose radius might take smaller values, with charge and bare mass
scaling with $R$ in such a way as to give sensible predictions. For example, both should 
vanish in the limit $R \rightarrow 0$. Therefore we propose
\be
q = \sigma R^n, \;\;\;\; m_0 = \rho R^{n'}
\ee
where $\sigma,\rho,n,n'$ are constants, with $\rho,n, n' > 0$. We also
assume that $\ffF_{\rm ext}$ varies with $R$ in a reasonable way. To be precise, we define
the 4-force per unit charge
\be
\tilde{\ffF}_{\rm ext}  \equiv \ffF_{\rm ext} / q
\ee
and we assume that $\tilde{\ffF}_{\rm ext}$ can expressed by a Taylor expansion about
$R=0$:
\be
\tilde{\ffF}_{\rm ext}(R) =  \sum_{k=0}^{\infty} \frac{ \tilde{\ffF}_{\rm ext}^{(k)}(0)}{k!} R^k .
\ee

The equation of motion is then
\begin{eqnarray}
&&
\sigma R^n \tilde{\ffF}_{\rm ext}(0) + \sigma \sum_{k=1}^{\infty} \frac{ \tilde{\ffF}_{\rm ext}^{(k)}(0)}{k!} R^{n+k}     \nonumber \\
&&+ \frac{2}{3} \sigma^2 \left(  - \dot{\ffU} R^{2n-1} + 
\left[ \ddot{\ffU} 
- \dot{\ffU}^2 \ffU \right] R^{2n}  + O(R^{2n+1}) \right)   \nonumber \\
&&= \rho R^{n'} \dot{\ffU}.
\label{motionRn}
\end{eqnarray}
We are at liberty to choose the powers $n,n'$ freely, as long as no unphysical or mathematically
dubious case is implied. Therefore we assume $n=n'$, which is natural---it means the shell
is an example of a class of shells whose charge to bare mass ratio is constant as the radius
varies (the assumption $n > n'$ is also viable; it merely complicates the analysis.) 
Eq. (\ref{motionRn}) has a well-behaved limit as $R \rightarrow 0$ as long as $2n-1 \ge 0$
so we assume
\[
n \ge 1/2.
\]
We can now find a solution curve by writing the Taylor series
\be
\ffU(R,\tau) = \sum_{k=0}^{\infty} \frac{ \ffU^{(k)}(0,\tau)}{k!} R^k .         \label{User}
\ee
We substitute this into  (\ref{motionRn}) and, in a first approximation, keep only the lowest order terms:
\be
\sigma R^n \tilde{\ffF}_{\rm ext}(0) =  \dot{\ffU}_0 \left[ \rho R^{n} + 
 \frac{2 \sigma^2}{3}  R^{2n-1} \right]           \label{eqlow}
\ee
where $\ffU_0 = \ffU(0,\tau)$. Note, this equation is accurate to order $R^\eta$ where $\eta = \min(n, 2n-1)$; the fact
that it contains some higher order terms when $n \ne 1$ does not change this.
An interesting case is $n=1$, which makes all these terms of the same order
and leads to the simplest analysis. However we don't require that $n=1$, only that $n \ge 1/2$.
Eq. (\ref{eqlow}) can be written
\be
q \tilde{\ffF}_{\rm ext}(0) =  m \dot{\ffU}_0        \label{eqmolow}
\ee
where
\be
m = m_0   + \frac{2 q^2}{3 R} = m_0 + m_{\rm ed} .    \label{mass}
\ee
Eq. (\ref{eqmolow}) is a well-behaved equation of motion.
It has precisely the form of the Lorentz force equation with no self-force, but with
the electrodynamics contributing to the observed mass. Eq. (\ref{mass}) warns us
that to model a `particle' of given charge and observed mass, we shall need to
assume $R \ge 2 q^2/3m$, and therefore the predictions cannot be expected to
be accurate if $\ffF_{\rm ext}$ or the worldline 
varies significantly on this distance or time scale. In particular, we require
\be
\ddot{\ffF}_{\rm ext} \ll \dot{\ffF}_{\rm ext} m / q^2.   \label{timescale}
\ee

The next set of terms in the power series (\ref{User}) is obtained by substituting (\ref{User})
into the equation of motion (\ref{motionRn}), retaining now all terms up to $O(R^{2n})$:
\be
\ffF_{\rm ext} + \frac{2}{3} q^2 \left(  \ddot{\ffU}_0 
- \dot{\ffU}_0^2 \ffU_0  \right) &=& m \dot{\ffU}  + O(R^{2n+1})  .            \label{ALDred}
\ee
By writing the full $\ffF_{\rm ext}$ on the left hand side, and the full $\dot\ffU$ on the
right hand side, we have also retained some terms of order higher than $R^{2n}$, but
this does not change the fact that this equation is valid to order $R^{2n}$.
Note that, in common with (\ref{eqmolow}), the equation can be written in terms 
of $q$, $m$ and $\ffF_{\rm ext}$  without explicitly mentioning $R$. The result
is an equation
of motion of the ALD form, but in which the self-force terms on the left hand side
are given by using the lowest order approximation to the worldline. In short, it is 
a `reduced order' ALD equation.

Since, in this method of solution, we already know $\ffU_0$ by the time we
attempt to solve (\ref{ALDred}), we may as well use the power series (\ref{User}) on
the right hand side and thus simplify the equation a little, obtaining
\be
\delta \ffF_{\rm ext}
+ \frac{2}{3} q^2 \left(  \ddot{\ffU}_0 
- \dot{\ffU}_0^2 \ffU_0  \right) &=&m \delta \dot{\ffU}  .
\ee
where we wrote $\delta\ffU \equiv \ffU - \ffU_0$ and 
$\delta \ffF_{\rm ext} \equiv \ffF_{\rm ext}(R) - \ffF_{\rm ext}(0)$.
This equation gives the perturbation in the worldline owing to the self-force
and the $R$-dependence of $\ffF_{\rm ext}$.  In this form one must propose a value
for $R$ or use some other strategy in order to obtain an expression for
$\delta \ffF_{\rm ext}$, but this is in any case necessary in order to evaluate
$\ffF_{\rm ext}$ if $\tilde{\ffF}_{\rm ext}$ depends on $R$.

Equation (\ref{ALDred}) is accurate to the order shown. However, it is convenient
to adjust the equation, while still retaining the same order of accuracy, so
as to make the left hand side orthogonal to the four-velocity. Then when one drops
those $O(R^{2n+1})$ terms that are not implicitly included, one has a well-constructed
four-vector equation, both sides of which are orthogonal to $\ffU$. To this end,
replace $\ddot{\ffU}_0$ by $\dot{\ffF}_{\rm ext}/m$ and
one occurrence of $\dot{\ffU}_0$ by $\ffF_{\rm ext}/m$, and elsewhere replace
$\ffU_0$ by $\ffU$:
\be
\ffF_{\rm ext} + \frac{2}{3} \frac{q^2}{m} \left(  \dot{\ffF}_{\rm ext} 
- \left(\dot{\ffU} \cdot \ffF_{\rm ext}\right) \ffU  \right) &=& m \dot{\ffU},
 \label{myEli}
\ee
where we now take it as understood that the equation is only accurate
to order $O(R^{2n})$. Taking the inner product with $\ffU$ one obtains
\be
\ffF_{\rm ext}\cdot \ffU + 
\frac{2}{3} \frac{q^2}{m} \left(  \dot{\ffF}_{\rm ext} \cdot \ffU
+ \dot{\ffU} \cdot \ffF_{\rm ext} \right) &=& 0.  
\ee
This is satisfied if the external force is pure, i.e. if
$\ffF_{\rm ext} \cdot \ffU = 0$ at all times (for example, this is
true for electromagnetic forces). For the case $\ffF_{\rm ext} = q \ftF \cdot \ffU$,
where $\ftF$ is the Faraday tensor of the applied electromagnetic field,
Eq. (\ref{myEli}) is Eliezer's equation (see appendix).
This justifies the use of the Eliezer or Ford-O'Connell equation in calculating the
motion of electrons and other charged particles. The Landau-Lifshitz
equation is almost identical, and may also be used, because it is accurate to
the same order of approximation---see appendix for details. 

The precision of the solution obtained by the above strategy is subject to
the considerations outlined in Gralla {\em et al.} \cite{09Gralla}. As time goes on, the departure from
the unperturbed worldline grows. Eventually the terms of any given order $R^n$ are no longer
small in comparison to those of lower order, and the approximations break down.
One method to deal with this is to start a new iteration of 
the calculation after some not too great elapse of proper time, and thus work ones way
along the worldline. 

In passing from (\ref{ALDred}) to (\ref{myEli}) the above treatment merely introduced replacements
in a such a way as to retain the order of approximation while passing to a form where
the higher order terms might possibly be zero. One could also obtain the relativistic Landau-Lifshitz
equation this way, so the present discussion does not offer any reason to prefer the equation
of Eliezer or Ford and O'Connell to that of Landau and Lifshitz. Ford and O'Connell have
given reasons to prefer their equation \cite{03OConnell,11Intravaia}. One cannot expect
either equation to be exact (within classical theory), because they do not account for details of the charge distribution
which one expects to influence the equation of motion at higher order, but for a specific entity
such as an electron there is a good chance that one equation is more accurate than the other.
Experimental exploration of this difference is challenging but may become possible
using modern laser techniques. The question can also be explored by numerical (as well
as analytical) treatments, since the whole framework of this problem remains fully within
ordinary classical electromagnetism with no adjustments needed; the only thing that
Maxwell's and Lorentz's equations cannot supply is a framework for calculating material
properties such as internal stresses (Poincar\'e stresses); these have to be supplied by assuming
an equation of state.

The ELL/FO equation is valuable because it is more accurate and more 
straightforward to work with than the ALD equation. The accuracy is better because
it accounts better for the charge distribution. It is also convenient because
there is less danger of making an unphysical parameter choice such as a too-small 
value of $R$ at given $q,m$, and because it is second not third
order (when regarded
as a differential equation for the 4-position), which makes it respect causality, 
The value of the above derivation (and others which arrive at the same conclusion) is
that it shows the ELL/FO equation should not be regarded as an approximation to the ALD equation;
rather, it is an alternative and better approximation to the exact equation of motion for a rigid
spherical charge. This is so because it is accurate to the same order in $R$ while avoiding the
properties which cause ALD to exhibit pathological
behavior. See \cite{11Bulanov,10Griffiths} for a comparison of their predictions for some example cases,
and the accuracy limit associated with the condition (\ref{timescale}).

To sum up, in this paper we have presented two main ideas. First we discussed
the absence of runaway behavior in the correct treatment of the motion of
charged bodies. The inertial term in the electromagnetic self force is often 
included implicitly rather than
explicitly in the equation of motion by absorbing it into the mass of the particles under
consideration. This can be correct, but only if one keeps in mind the need
to avoid unphysical parameter values, such as negative total rest energy in some region of space. 
In all cases that have been calculated, and total self-force of an accelerated
object {\em opposes} the acceleration. Therefore there is no known physical object, having
a positive bare mass, whose equation of motion has runaway solutions in the absence of
an applied force, when that motion is treated by standard classical electromagnetism
(Maxwell's equations and the Lorentz force equation).

The second main idea of this paper was to present an approach to handling the 
ALD equation, by considering it to be what it is---an approximate equation of motion
of a rigid charged shell, not an exact equation for point particles. 
We have given a straightforward method to form the $R \rightarrow 0$ limit 
to the equation of motion in a physically and mathematically sensible manner.
This is similar to the approach taken by Gralla {\em et al.}, but considerably simpler.
The result is a reduced-order ALD equation ((\ref{ALDred}) or (\ref{myEli})) previously
recommended by several other authors.

I thank V. Hnizdo, S. Deser and R. F. O'Connell for helpful comments.

\appendix
\section{Landau-Lifschitz equation}

Eq. (53) of Eliezer \cite{48Eliezer} is, in our notation,
\be
m \dot{\ffU} = q \ftF \cdot \ffU + \frac{2}{3} \frac{q^3}{m} \left(
\dby{}{\tau} \left( \ftF \cdot \ffU \right) - \ffU (\ftF \cdot \ffU) \cdot \dot{\ffU} \right).
\label{Eli}
\ee
Eq. (12) of Gralla {\em et al.} \cite{09Gralla} is, in our notation,
\be
m \dot{\ffU} = q \ftF \cdot \ffU + \frac{2}{3} \frac{q^3}{m} \left( \ffw + \ffU (\ffU \cdot \ffw) \right)
\ee
where
\be
\ffw = \dot{ \ftF } \cdot \ffU + \frac{q}{m} \ftF \cdot (\ftF \cdot \ffU).  \label{wdef}
\ee
These equations are very similar but not identical. Since $\ftF$ is antisymmetric, we
have $\ffU \cdot \ftF \cdot \ffU = 0$ and therefore
\be
\dot{\ffU}\cdot(\ftF\cdot\ffU) = - \ffU \cdot \dby{}{\tau}\left(\ftF\cdot\ffU \right).
\ee
Substituting this into (\ref{Eli}), Eliezer's equation can be written
\be
m \dot{\ffU} = q \ftF \cdot \ffU + \frac{2}{3} \frac{q^3}{m} \left( \tilde{\ffw} + \ffU (\ffU \cdot \tilde{\ffw})
\right)            \label{Eliw}
\ee
where
\be
\tilde{\ffw} = \dby{}{\tau} \left( \ftF \cdot \ffU \right) = \dot{\ftF}\cdot \ffU + \ftF \cdot \dot{\ffU}.
\ee
Comparing this with (\ref{wdef}), we see that, to convert the Eliezer equation into the 
Landau-Lifshitz equation one may replace the acceleration $\dot{\ffU}$ in the expression for
$\tilde{\ffw}$ by its value ignoring self force, i.e. $(q/m) \ftF \cdot \ffU$. This results in
an equation that is valid
to the same order of approximation, and easier to work with. As long
as one does not make the mistake of assuming either equation is exact, either may be used.

%\begin{thebibliography}{99}
%\end{thebibliography}

\bibliography{selfforcerefs}

\end{document}